\documentclass[onecolumn]{aastex62}

\usepackage{lineno}
\usepackage{longtable}

\graphicspath{./}

\begin{document}

\title{Observational Properties of 155 $\beta$ Cephei pulsating variable stars}

\correspondingauthor{Xiang-dong Shi}
\email{sxd@ynao.ac.cn}

\author{Xiang-dong Shi}
\affiliation{Yunnan Observatories, Chinese Academy of Sciences(CAS), P.O. Box 110, Kunming 650216, P.R. China}
\affiliation{Department of Astronomy, Key Laboratory of Astroparticle Physics of Yunnan Province, Yunnan University, Kunming 650091, P.R. China}
\affiliation{University of Chinese Academy of Sciences, No.1 Yanqihu East Rd, Huairou District, Beijing 101408, P.R. China}
\affiliation{Key Laboratory of the Structure and Evolution of Celestial Objects, CAS, Kunming 650216, P.R. China}

\author{Sheng-bang Qian}
\affiliation{Department of Astronomy, Key Laboratory of Astroparticle Physics of Yunnan Province, Yunnan University, Kunming 650091, P.R. China}
\affiliation{University of Chinese Academy of Sciences, No.1 Yanqihu East Rd, Huairou District, Beijing 101408, P.R. China}
\affiliation{Key Laboratory of the Structure and Evolution of Celestial Objects, CAS, Kunming 650216, P.R. China}

\author{Li-ying Zhu}
\affiliation{Yunnan Observatories, Chinese Academy of Sciences(CAS), P.O. Box 110, Kunming 650216, P.R. China}
\affiliation{University of Chinese Academy of Sciences, No.1 Yanqihu East Rd, Huairou District, Beijing 101408, P.R. China}
\affiliation{Key Laboratory of the Structure and Evolution of Celestial Objects, CAS, Kunming 650216, P.R. China}

\author{Lin-jia Li}
\affiliation{Yunnan Observatories, Chinese Academy of Sciences(CAS), P.O. Box 110, Kunming 650216, P.R. China}
\affiliation{University of Chinese Academy of Sciences, No.1 Yanqihu East Rd, Huairou District, Beijing 101408, P.R. China}
\affiliation{Key Laboratory of the Structure and Evolution of Celestial Objects, CAS, Kunming 650216, P.R. China}

\author{Er-gang Zhao}
\affiliation{Yunnan Observatories, Chinese Academy of Sciences(CAS), P.O. Box 110, Kunming 650216, P.R. China}
\affiliation{University of Chinese Academy of Sciences, No.1 Yanqihu East Rd, Huairou District, Beijing 101408, P.R. China}
\affiliation{Key Laboratory of the Structure and Evolution of Celestial Objects, CAS, Kunming 650216, P.R. China}

\author{Wen-xu Lin}
\affiliation{Yunnan Observatories, Chinese Academy of Sciences(CAS), P.O. Box 110, Kunming 650216, P.R. China}
\affiliation{University of Chinese Academy of Sciences, No.1 Yanqihu East Rd, Huairou District, Beijing 101408, P.R. China}
\affiliation{Key Laboratory of the Structure and Evolution of Celestial Objects, CAS, Kunming 650216, P.R. China}

\begin{abstract}
$\beta$ Cephei pulsating variable (BCEP) stars are the most massive pulsating variable stars in the main sequence, exhibiting both p- and g-mode pulsations.
In this study, we identified 155 BCEP stars or candidates using data from TESS and Gaia, of which 83 were first confirmed as BCEP stars. They have visual magnitudes ranging from 8 to 12 mag and effective temperatures between approximately 20,000 and 30,000 K, while the parallaxes of most targets are between 0.2 and 0.6 mas. The study indicates that these BCEP stars have pulsation periods ranging from 0.06 to 0.31 days, with amplitudes ranging from 0.1 to 55.8 mmag in the TESS band. Additionally, the number of BCEP stars increases as the pulsation amplitude decreases. These targets align with the distribution region of BCEP stars in the luminosity-period (L-P) and temperature-period (T-P) diagrams. We have updated the L-P relation of BCEP stars. The Hertzsprung-Russell (H-R) diagram indicates that these targets are in the main-sequence evolutionary phase, with masses ranging from 7 to 20 $M_{\odot}$ and luminosities between 2800 and 71,000 $L_{\odot}$. They are almost in the theoretical instability region of BCEP stars but as previously reported, this region at the low-mass end (red) is not filled. The distribution of the pulsation constant indicates that the dominant pulsation periods of BCEP stars consist mainly of low-order p-mode pulsations with a high proportion of radial fundamental modes. These BCEP stars are excellent objects for enhancing our understanding of the structure and evolution of massive stars through asteroseismology.
\end{abstract}


\section{Introduction} \label{sec:intro}

Massive stars usually refer to O- and B-type (OB-type) stars, with properties of high mass, temperature, and luminosity \citep{1968ApJ...151..611M, 1973ARA&A..11...29M, 1973AJ.....78..929P}. Their properties and evolutions correlate with many important physical processes and objects, such as supernova explosions, gravitational-wave events, neutron stars, black holes, and so on \citep{2010ApJ...725..940Y, 2008ApJ...676.1162S, 2020RAA....20..161H, 2020A&A...638A..39L}. Among massive stars, pulsating objects are very important, because their internal structure information can be obtained through the asteroseismology method (e.g., \citet{2003Sci...300.1926A, 2006A&A...459..589M, 2006ApJ...642..470A, 2006MNRAS.365..327H, 2008MNRAS.385.2061D, 2012MNRAS.427..483B}).

The massive pulsating stars in the main-sequence (MS) evolution stage mainly include $\beta$ Cephei pulsating variable (BCEP) stars and slowly pulsating B-type (SPB) stars. They show variations in line profiles and luminosity \citep{2002ASPC..259..196D}. Both BCEP and SPB stars are excited by the $\kappa$ mechanism \citep{1992A&A...256L...5M, 1993MNRAS.262..204D}, which operates in the ionization zone of iron-group elements. Among them, BCEP stars have a spectral type ranging from late O-type to early B-type \citep{2007CoAst.151...48M} and typically exhibit periods ranging from 2 to 7 hours \citep{2005ApJS..158..193S}. BCEP stars pulsate in low-order p- and g-mode \citep{2005ApJS..158..193S}, making them excellent candidates for detailed asteroseismic investigations. This is because the physical properties of the propagation region determine the frequency of each oscillation mode, and measuring these frequencies can constrain the physical conditions in the stellar interior (e.g., \citet{2003Sci...300.1926A, 2006A&A...459..589M, 2006ApJ...642..470A, 2006MNRAS.365..327H, 2008MNRAS.385.2061D, 2012MNRAS.427..483B}).

The photometric and spectroscopic variability of the $\beta$ Cephei prototype star was discovered by \citet{1902ApJ....15..340F} over 100 years. However, the pulsation of BCEP stars, excited by the $\kappa$ mechanism, was not discovered until 90 years later \citep{1992A&A...256L...5M, 1993MNRAS.262..204D}. \citet{2005ApJS..158..193S} provided an excellent overview of the observational features of BCEP stars and compiled a catalog of 93 confirmed BCEP stars based on data from papers spanning nearly 100 years. \citet{2008A&A...477..917P} found 103 new BCEP stars based on the ASAS-3 data. Three new BCEP stars were identified by \citet{2019MNRAS.489.1304B} using data from the \emph{K2} mission. \citet{2020AJ....160...32L} detected 113 BCEP stars (including 86 new objects) from the data of the KELT exoplanet survey. \citet{2020MNRAS.493.5871B} identified 327 BCEP stars using data from the Transiting Exoplanet Survey Satellite (TESS) in sectors 1-18. After cross-matching these catalogs and removing duplicate targets, we concluded that the number of BCEP stars discovered in our Galaxy so far is about 400. However, due to the small sample size, many features of BCEP stars may not be well understood. Therefore, it is essential to search for more samples. Especially due to the low-amplitude variation of BCEP stars, systematic high-precision space surveys will provide us with an unprecedented opportunity to search for such targets.

The $Gaia$ Satellite was launched on December 19, 2013, as a European space mission \citep{2016A&A...595A...1G, 2018A&A...616A...1G, 2021A&A...649A...1G}. It was designed to collect astrometry, photometry, and spectroscopy data for nearly 2 billion objects in the Milky Way, extragalactic systems, and the solar system. The stellar luminosity can be estimated based on the parallax measurements of Gaia, which provide an independent and crucial method for studying variables or assessing the reliability of our results (e.g., \citet{2021PASP..133e4201S}). Meanwhile, the recently released Gaia DR3 data also provides the stellar atmospheric parameters from the low-resolution spectral data of the Blue and Red Photometer, as well as the spectrum from the Radial Velocity Spectrometer. These parameters are derived by different models, such as the General Stellar Parametrizer from Photometry (GSP-Phot) and the Extended Stellar Parametrizer for Hot Stars (ESP-HS).

The TESS was launched by NASA in April 2018 \citep{2015JATIS...1a4003R}. It can simultaneously monitor a 24$\times$96 degree area of the sky using four wide-field cameras. The sky is divided into 26 strips, each of which is observed continuously for 27 days, allowing the entire sky to be scanned once every two years. The primary mission of TESS is to search for transiting exoplanets across the entire sky. Moreover, its high-precision and extensive photometric data also provide a unique opportunity to systematically study variable stars across the entire sky, particularly focusing on bright stars (e.g., \citet{2020MNRAS.493.5871B, 2022ApJS..259..50S}), such as those objects with visual magnitudes less than 15, which depends on the characteristics of stars and our scientific goals. Recently, an increasing amount of research on variable stars using TESS data has been published (e.g., \citet{2023ApJS..265...33S, 2023ApJS..268...16S}, hereafter referred to as Papers I and II), highlighting its significance in the study of variable stars.

In this study, 155 BCEP stars are identified using observation data from TESS and Gaia, and their observational and physical characteristics are further analyzed. On one hand, we analyzed the light curves of these objects observed by TESS and obtained their photometric characteristics. On the other hand, we calculated their luminosities based on their parallaxes, temperatures, and extinctions provided by Gaia, and conducted statistical analysis of their physical characteristics.

\section{$\beta$ Cephei pulsating variable stars observed by TESS and GAIA} \label{}

We acquired the TESS photometric data with a 2-minute short cadence for sectors 1-55 from the Mikulski Space Telescope Archive (MAST) database. To identify new targets, we first excluded the BCEP stars listed in the General Catalog of Variable Stars (GCVS) \citep{2017ARep...61...80S}. Then, we applied the methods outlined in \citet{2021AJ....161...46S, 2021MNRAS.505.6166S} to process the data as light curves, similar to differential photometry. This involved converting the flux to magnitude and subtracting the average magnitude. Here, we chose to utilize the Pre-search Data Conditioned Simple Aperture Photometry (PDCSAP) data, which has been processed to remove instrumental variations and excessive scattered light \citep{2010ApJ...713L..87J}.

Based on the definition and properties of BCEP stars provided by \citet{2005ApJS..158..193S}, we identified 155 BCEP stars (including 4 candidates) using a straightforward program-assisted visual classification from the light curves of TESS. The information about these 155 objects is provided in Table \ref{tab:1}. The purpose of this straightforward program is to filter out targets that only have frequencies with a signal-to-noise ratio (S/N) less than a specified threshold. The next section will provide the definition and threshold of S/N. The criteria for visual classification include observable features such as their position in the H-R diagram, the presence of periodic signals that are too short to be binary variation or rotational modulation, and the Fourier spectral properties of BCEP stars, which will be specifically discussed in Section 3.

We compared the identified samples with the published work presented in Section 1 and found that 76 objects identified as BCEP stars or candidates are listed in the work of \citet{2020AJ....160...32L} or \citet{2020MNRAS.493.5871B}. In addition, 33 samples were also listed as MS oscillators by Gaia DR3. The details are listed in the last Column of Table \ref{tab:1}. There are no samples in the list of \citet{2005ApJS..158..193S, 2008A&A...477..917P}, because the targets in these lists are included in GCVS and are already being removed. Among these 155 objects, 83 of them are confirmed as BCEP stars for the first time.

\begin{figure*}\centering \vbox to3.0in{\rule{0pt}{5.0in}}
\includegraphics{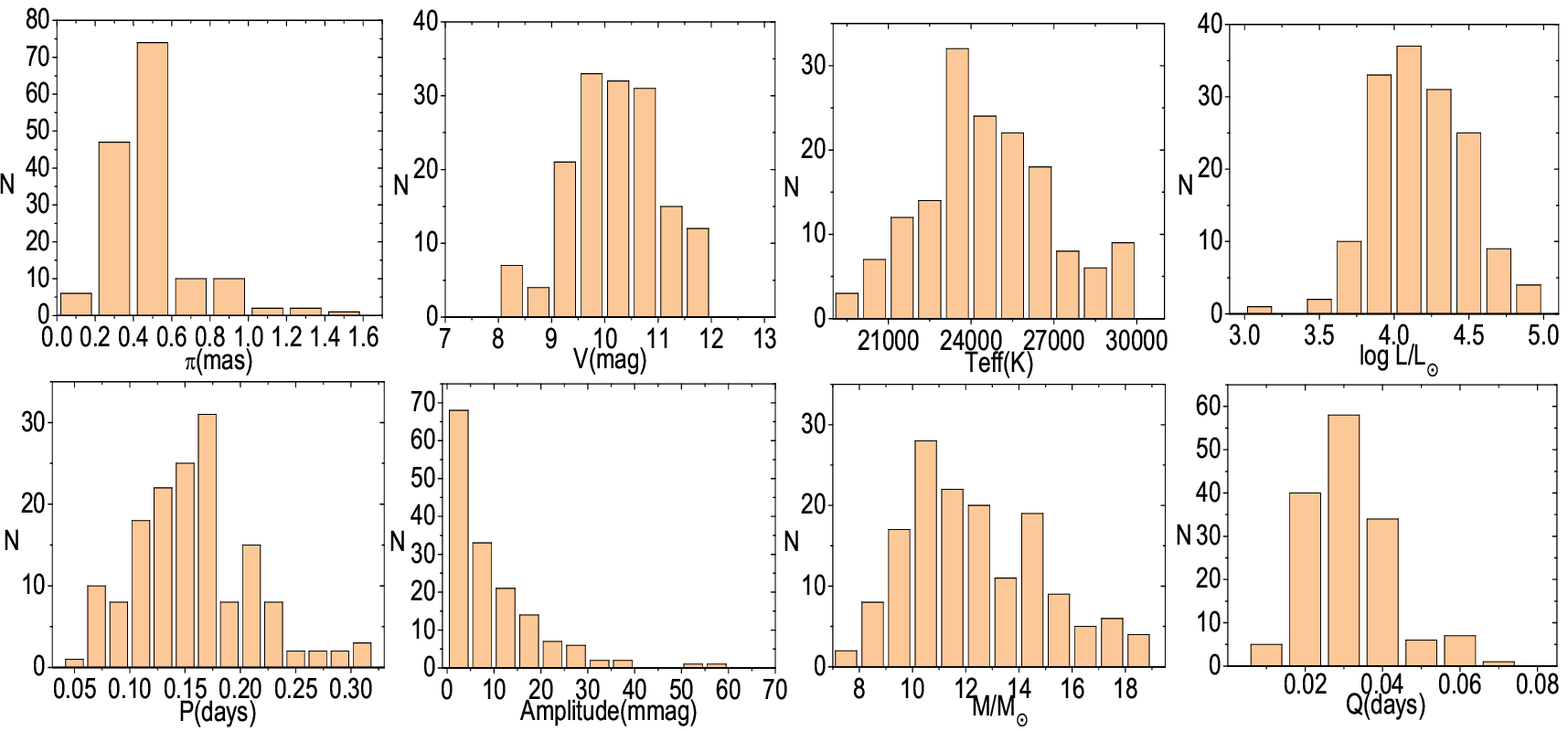}
\caption{Distribution of the parameters for BCEP stars.}
\label{fig:distribution}
\end{figure*}

The parallaxes $\pi$ of these samples are measured by the Gaia Survey and are shown in Column (2) of Table \ref{tab:1}. The parallax data from Gaia DR3 is used for the first priority, while the results from Gaia DR2 are used for several targets that exhibited outliers. Nevertheless, three targets (TIC29585482
, 246785266, and 216167514) either lack parallax values or have significantly unreasonable negative values that should be excluded. Figure \ref{fig:distribution} illustrates the parallax distribution of these BCEP stars, revealing that the parallax of most targets is between 0.2 and 0.6 mas. Column (3) of Table \ref{tab:1} provides the visual magnitude $m_{V}$ from the TESS Input Catalog (TIC, \citet{2018AJ....156..102S}) or SIMBAD \citep{2000A&AS..143....9W}. The $m_{V}$ distribution of these targets is also displayed in Figure \ref{fig:distribution}, indicating that these BCEP stars have visual magnitudes between 8 and 12 mag.

The resolution of TESS is 21 arcseconds per pixel, so it is necessary to check for contamination from neighboring stars within the photometric aperture. We used SIMBAD and Gaia to verify the presence of stars with a similar brightness to our target within the 1 arc minute field of view surrounding our target. If there are any contaminants, the target should be marked with the capital letter "C" in parentheses in Column (10) of Table \ref{tab:1}.

As demonstrated in Papers I and II, Gaia ESP-HS reliably provides temperature measurements for hot stars in the majority of cases. Therefore, the temperatures of these BCEP stars in this study are all from Gaia ESP-HS and are listed in Table 1. Other atmospheric parameters will not be used in this study due to the uncertainty of their reliability.

\begin{figure*}\centering \vbox to3.0in{\rule{0pt}{5.0in}}
\includegraphics{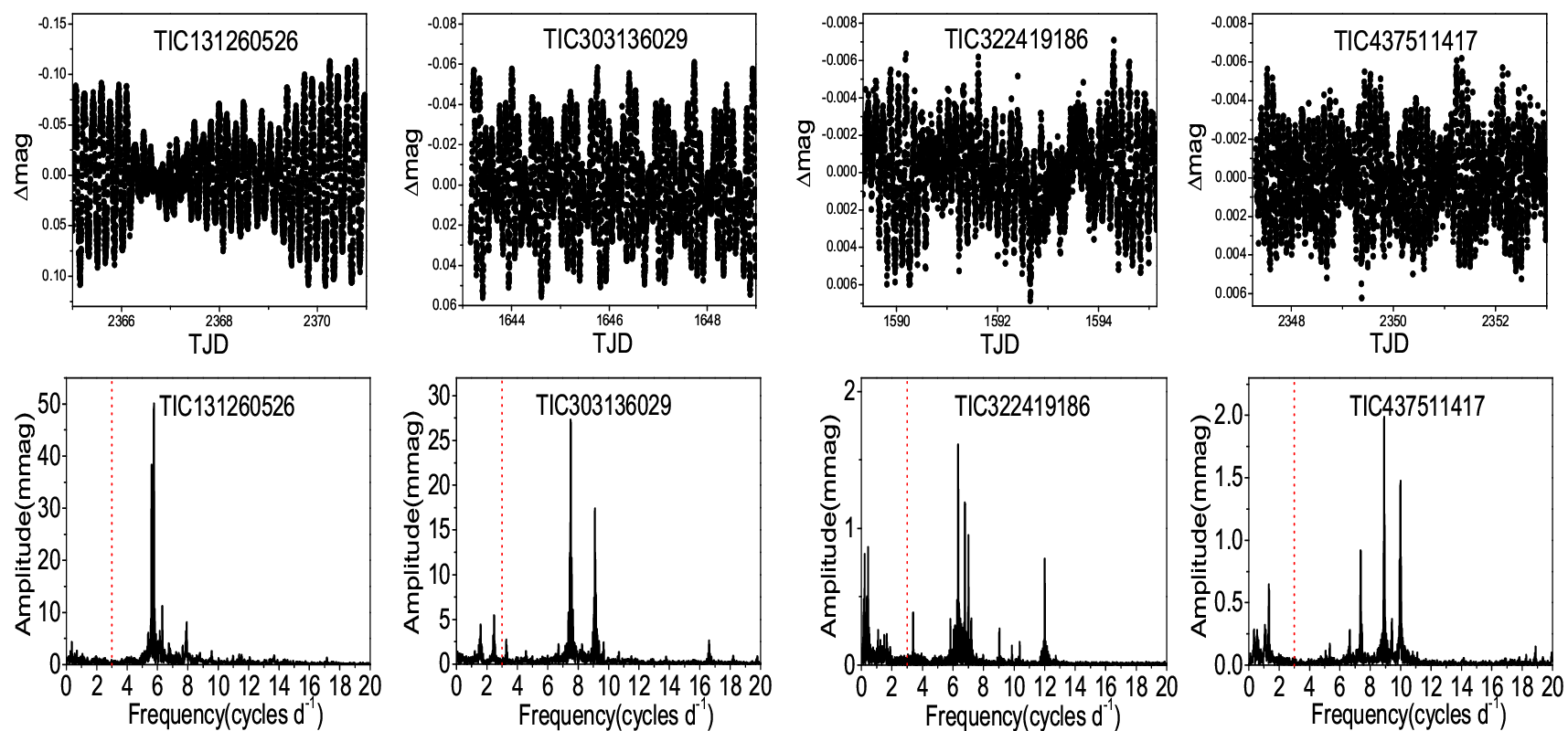}
\caption{Example light curves and Fourier spectra for some BCEP stars. The red dotted lines represent the approximate boundary 3 cycles $d^{-1}$ for low-frequency and high-frequency. The Fourier spectra of 83 first confirmed BCEP stars are shown in the appendix.}
\label{fig:1}
\end{figure*}

\begin{figure*}\centering \vbox to4.0in{\rule{0pt}{5.0in}}
\includegraphics{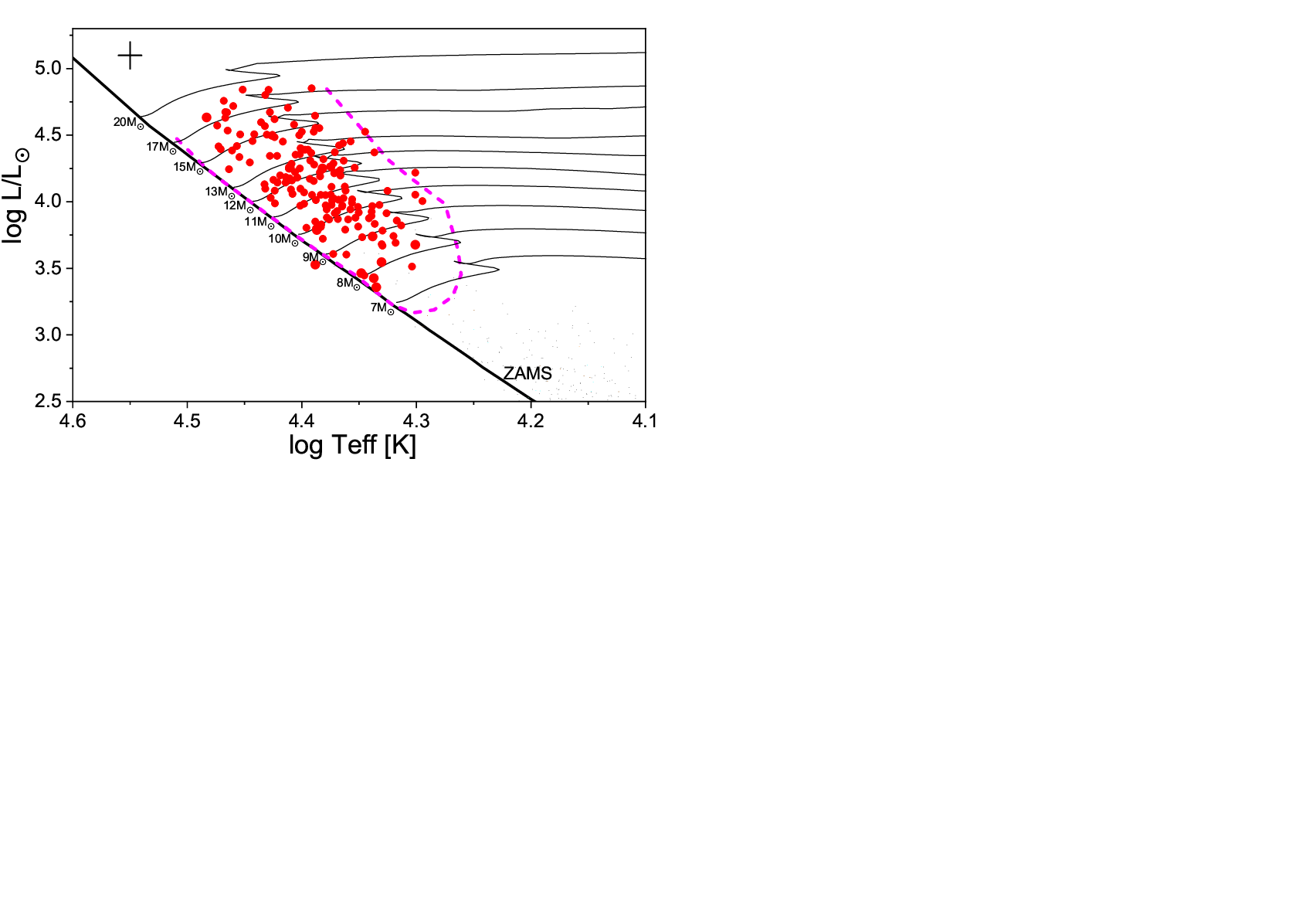}
\caption{The H-R diagram of these BCEP stars. The red solid circles represent these BCEP stars and candidates. The BCEP and SPB stars, which were previously published in Papers I and II, are also depicted as red open circles and black solid circles, respectively. Targets with suspected unreliable parameters in Papers I and II were excluded from this diagram. The theoretical zero-age main sequence and evolutionary paths for various masses with Z=0.02 are represented by the black solid lines. The blue and magenta dotted lines represent the theoretical instability regions of SPB and BCEP stars for Z = 0.02 and spherical harmonic degree $l \leq$ 3, as documented in \citet{2007CoAst.151...48M}. The black cross in the upper left corner represents a standard error box.}
\label{fig:L-T}
\end{figure*}

\begin{figure*}\centering \vbox to4.0in{\rule{0pt}{5.0in}}
\includegraphics{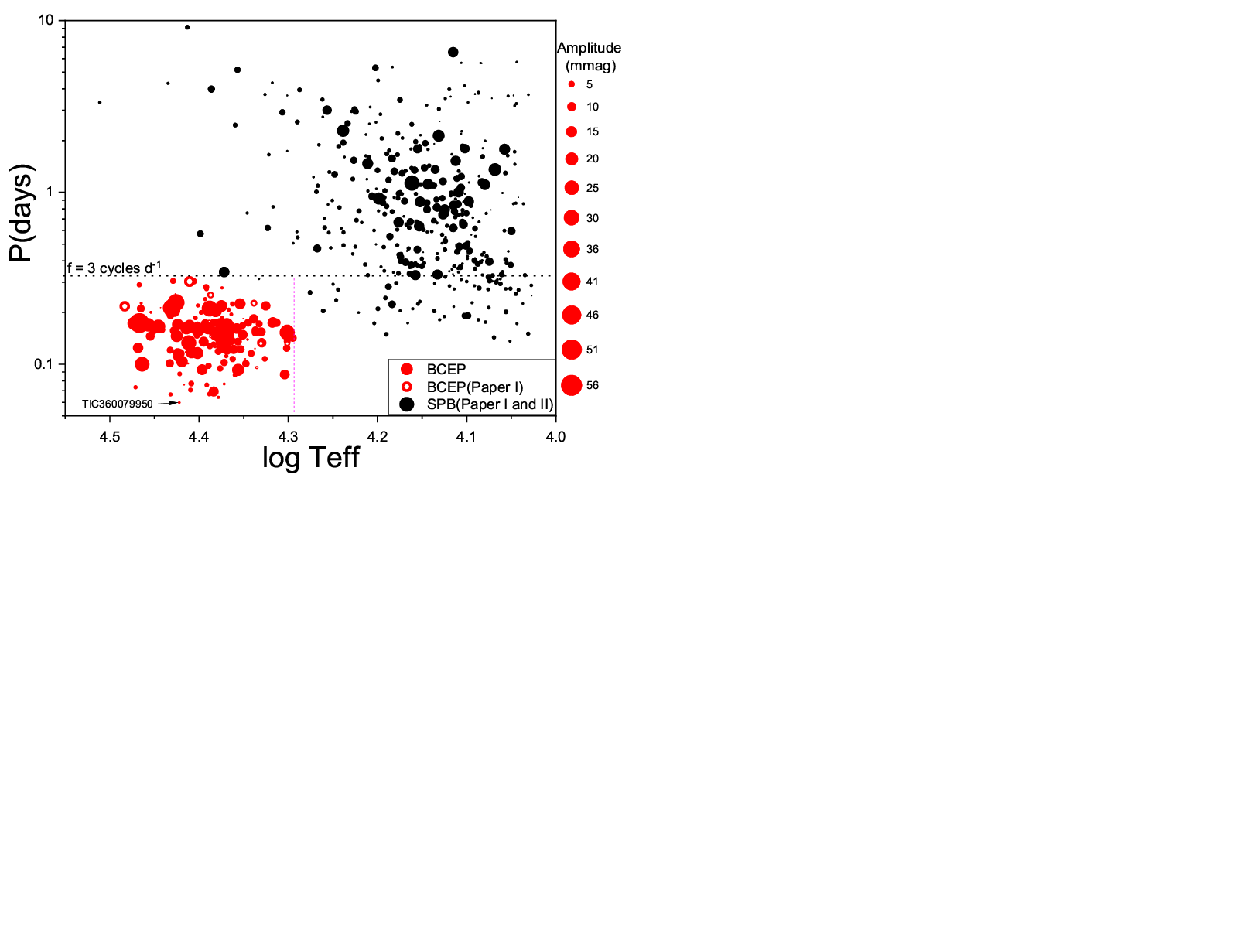}
\caption{The effective temperature and the dominant pulsating period relation diagram of these BCEP stars. Similar symbols to those in Figure \ref{fig:L-T} are used, but the size of the circles denotes their pulsation amplitude at the dominant frequency.}
\label{fig:T-P}
\end{figure*}

\begin{figure*}\centering \vbox to4.0in{\rule{0pt}{5.0in}}
\includegraphics{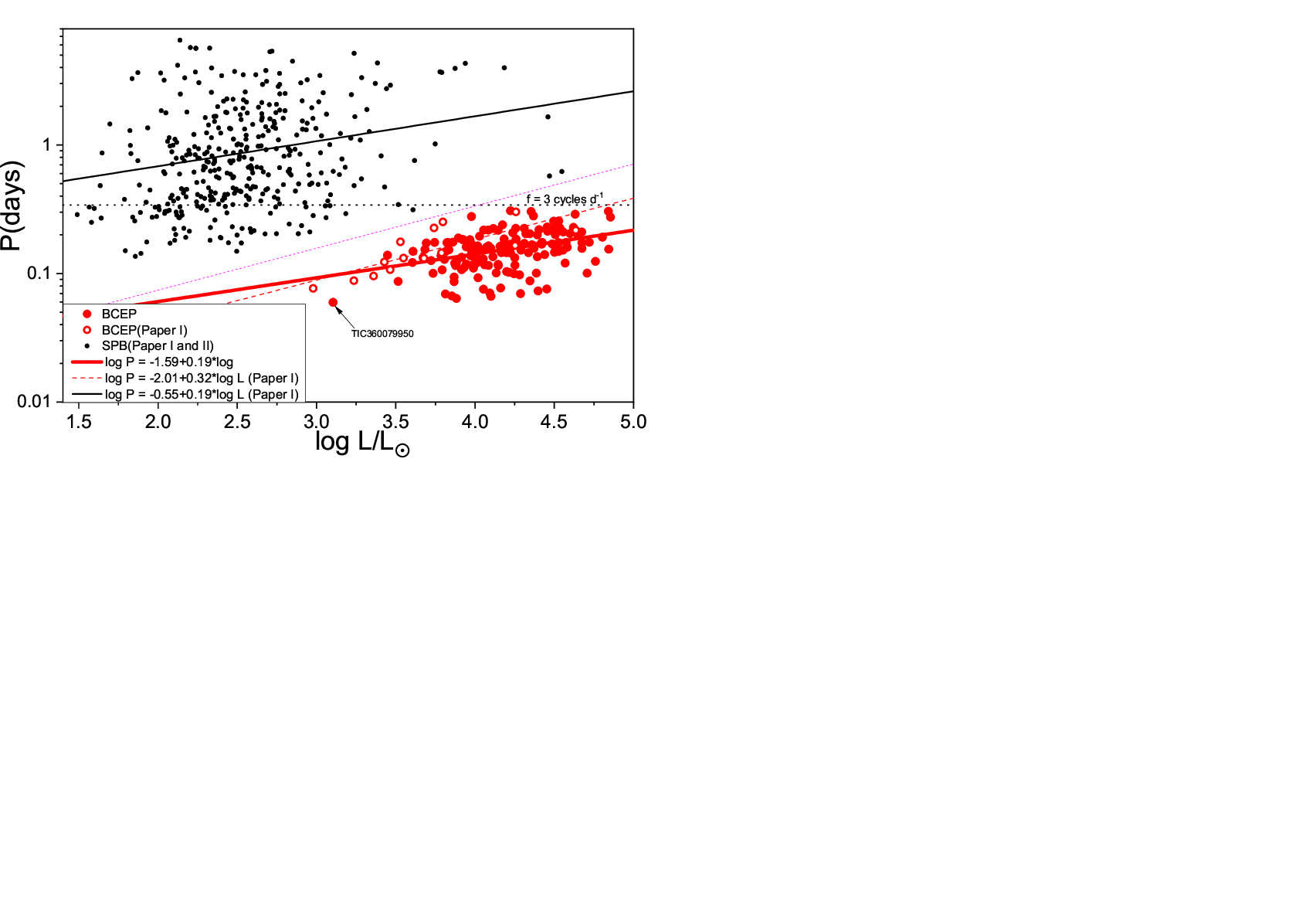}
\caption{The luminosity and the dominant pulsating period relation diagram of these BCEP stars. Symbols are similar to those in Figure \ref{fig:L-T}. The red dashed line and the black solid line represent the relationship derived by Paper I for BCEP and SPB stars, respectively. The red solid line represents the new relation derived in this paper.}
\label{fig:L-P}
\end{figure*}

\section{The properties of these $\beta$ Cephei pulsating variable stars}

\subsection{Light curves and Fourier spectra}

Calculating the Fourier spectrum is a common method for studying the luminosity variations of pulsating stars. We obtained the Fourier spectra of the light curves for these BCEP stars using the Period04 software \citep{2005CoAst.146...53L}. Figure \ref{fig:1} shows some examples of the light curves and Fourier spectra for these BCEP stars. The Fourier spectra of 83 first confirmed as BCEP stars are shown in the Appendix. The signal-to-noise ratio (S/N) is the ratio of signal amplitude to the average noise amplitude near the frequency. It is the criterion used to determine the reliability of a detected frequency, which is calculated using a box size of 1 $d^{-1}$ and the residuals at the original, and a threshold of S/N $\geq$ 5.4 is adopted for a 2-minute short cadence \citep{2021AcA....71..113B}.

Based on the period range of p-mode pulsation less than 0.3 days(e.g., \citet{2010aste.book.....A, 2005ApJS..158..193S}), the value of 3 cycles $d^{-1}$ is assumed as a rough boundary between low-frequency and high-frequency in the Fourier spectrum. The frequencies greater than 3 cycles $d^{-1}$ cannot be a binary variation or rotational modulation, because main sequence stars with a mass greater than 7 $M_{\odot}$ typically have a radius greater than 4 $R_{\odot}$, corresponding to an equatorial rotation speed greater than 600 $km \cdot s^{-1}$.

In Columns (8) and (9) of Table \ref{tab:1}, we list the periods and amplitudes corresponding to their dominant frequencies, along with the uncertainty on the last one bit in parentheses, as derived from \citet{1999DSSN...13...28M}. The distributions of periods and amplitudes are displayed in Figure \ref{fig:distribution}. The figure illustrates that these BCEP stars exhibit pulsation periods ranging from 0.06 to 0.31 days with a peak at approximately 0.17 days, and pulsation amplitudes ranging from 0.1 to 55.8 mmag. The number of BCEP stars increases as the amplitude decreases.

Column (10) of Table \ref{tab:1} displays the recommended variable star type. A prerequisite for assessing BCEP stars is the existence of p-mode pulsation. In terms of observable measurements, there should be at least one high frequency exceeding 3 cycles $d^{-1}$. However, certain harmonic or combined frequencies of the g-mode also satisfy this condition (e.g., \citet{2015MNRAS.450.3015K}, Papers I and II ), and therefore they should be excluded. We labeled the targets with relatively certain p-mode pulsation as "BCEP" and those in an eclipsing binary (EB) system as "EB+BCEP". There are four targets with limited frequency resolution, making it difficult to conclude whether they exhibit p-mode pulsations, harmonic, or combined frequencies of the g-mode. However, they are suspected to be BCEP stars and are marked as "BCEP?".

\subsection{Luminosity and H-R diagram}

Except for the three targets without parallax, Column (7) of Table \ref{tab:1} provides the luminosities of these BCEP stars, derived using the following formulae,
\begin{equation}
log(L_/L_{\odot})=0.4\times{(4.74-M_{V}-BC)}
\end{equation}
\begin{equation}
M_{V}=m_{V}-5\times{log(1000/\pi)}+5-A_{V},
\end{equation}
where the interstellar extinction $A_{V}$ is from Gaia DR3, and the parallax $\pi$ and visual magnitude $m_{V}$ are as described in Section 2. The bolometric correction $BC$ is a function of temperature given by \citet{2020MNRAS.495.2738P} for the JOHNSON.\emph{V} passband using the grids that combine both local thermodynamic equilibrium (LTE) and non-local thermodynamic equilibrium (NLTE). Based on the typical errors of 0.05 mas, 0.10 mag, 0.02 mag and 0.01 mag for $\pi$, $A_{V}$, $BC$ and $m_{V}$ respectively, the mean uncertainties of $log(L_/L_{\odot})$ are estimated to be around 0.1 dex (e.g., \citet{2020MNRAS.493.5871B}).

The luminosity distributions are depicted in Figure \ref{fig:distribution}, revealing that these BCEP stars, with the exception of TIC360079950, have luminosities ranging from 2800 to 71000 $L_{\odot}$. The parallax of TIC360079950 given by Gaia DR3 is 1.46 mas, which is the highest among these BCEP stars. The value given by Gaia DR2 is -5.75 mas, which is an anomalous value. The Gaia DR3 RUWE value for TIC 360079950 is 10.965, indicating that it is likely a binary system. Therefore, we believe that the 1.46 mas parallax of TIC360079950 is unreliable, possibly due to the orbital motion of binary system, and its extremely low luminosity is likely caused by the unreliable parallax. Of course, it may only be a subdwarf star in a binary system, which seems to be supported by the fact that it has the highest frequencies in these sample.

By combining the temperatures, an H-R diagram can be constructed in Figure \ref{fig:L-T}, illustrating these BCEP stars along with the SPB and BCEP stars previously published in papers I and II. In this figure, we used the stellar evolution code Modules for Experiments in Stellar Astrophysics (MESA, \citet{2011ApJS..192....3P, 2013ApJS..208....4P, 2015ApJS..220...15P, 2018ApJS..234...34P, 2019ApJS..243...10P} to create the theoretical Zero Age Main Sequence (ZAMS) and the evolutionary tracks of stars with different masses, for Z = 0.02. This figure also illustrates the theoretical instability regions for SPB and BCEP stars with a spherical harmonic degree $l \leq$ 3 and Z = 0.02 as presented in \citet{2007CoAst.151...48M}.

The H-R diagram indicates that, with the exception of TIC360079950, these BCEP stars are in the main sequence evolutionary phase with masses ranging from 7 to 20 $M_{\odot}$. It can also be observed in the H-R diagram that these BCEP stars are distributed within the theoretical instability region of BCEP stars. However, the theoretical instability region of BCEP stars at the low-mass end (red) remains unfilled.

\subsection{Masses and Pulsation constants}

By comparing the positions of stars in H-R diagram with the evolutionary tracks, we estimated the mass of these BCEP stars. This information can also be used to calculate the pulsation constant Q, which corresponds to the dominant pulsation frequency. The error of Q is mainly caused by T and L, estimated to be around 30\% (e.g., \citet{2005ApJS..158..193S}).

Except for the four targets that did not obtain reliable luminosity, the mass distribution of these BCEP stars is shown in Figure 1. Most of these targets have masses between 8 and 16 $M_{\odot}$, with a peak around 11 $M_{\odot}$. Figure 1 also illustrates the distribution of pulsation constants for these targets, indicating that most targets are distributed between 0.015 and 0.045 days, with a peak around 0.033 days.

\subsection{The T-P and L-P diagram}

As described in Papers I and II, in addition to the H-R diagram, the T-P and L-P diagrams are very helpful in distinguishing between BCEP and SPB stars from various perspectives. This is because BCEP and SPB stars have overlapping regions in the H-R diagram, but there is almost no overlap in the T-P and L-P diagrams. In addition, T-P or L-P diagrams can also help identify targets with unreliable temperature or luminosity measurements, such as TIC360079950.

Figures \ref{fig:T-P} and \ref{fig:L-P} depict the T-P and L-P diagrams of these new BCEP stars, as well as the SPB and BCEP stars previously published in Papers I and II. These BCEP stars are located in a region that is distinct from the region of SPB stars on the T-P and L-P diagrams. Meanwhile, we have also utilized the least squares method to update the L-P relation for BCEP stars as
\begin{equation}\label{eq:L-P}
log P = -1.59(\pm0.15)+0.19(\pm0.04)\times{log L}.
\end{equation}

\section{Discussion and conclusion}

We identified 155 BCEP stars or candidates using the photometric data from TESS and the spectral and astrometric data from Gaia. Among these targets, 83 stars are confirmed for the first time as BCEP stars. These BCEP stars have visual magnitudes ranging from 8 to 12 mag and effective temperatures ranging from about 20,000 to 30,000 K. The parallaxes of most targets are between 0.2 and 0.6 mas. After analyzing the Fourier spectra of the light curves for these targets, we determined their dominant pulsation periods and amplitudes. These BCEP stars have pulsation periods ranging from 0.06 to 0.31 days, with a peak at approximately 0.17 days, which is in close agreement with the finding of \citet{2005ApJS..158..193S}. The amplitude distribution in the TESS band is ranging from 0.1 to 55.8 mmag, and the number of BCEP stars increases as the amplitude decreases.

Meanwhile, we estimated the masses and pulsation constants Q for these BCEP stars. The masses of most targets are between 8 and 16 $M_{\odot}$, with a peak around 11 $M_{\odot}$, which is consistent with the results of \citet{2005ApJS..158..193S}.
The distribution of pulsation constant Q for these targets indicates that most targets are distributed between 0.15 and 0.45 days, with a peak around 0.033 days, which should correspond to the radial fundamental mode pulsation. This indicates that the dominant pulsation periods of BCEP stars consist mainly of low-order p-mode pulsations with a high proportion of radial fundamental modes.

The H-R diagram indicates that nearly all of these stars are in the main sequence evolutionary phase with masses ranging from 7 to 20 $M_{\odot}$ and luminosities between 2800 and 71,000 $L_{\odot}$. They are almost in the theoretical instability region of BCEP stars. The theoretical instability regions of BCEP stars at the low-mass end (red) have not been filled, as reported by \citet{2005ApJS..158..193S}. However, in contrast to the results of \citet{2005ApJS..158..193S}, the theoretical instability region of BCEP stars is filled for masses ranging from 16 to 20 $M_{\odot}$.

These BCEP stars are located in a region that is distinct from the region of SPB stars on the T-P or L-P diagrams. We have also updated the L-P relation for BCEP stars using the least squares method. The new L-P relation indicates that the slope of BCEP stars is consistent with that of SPB stars within the margin of error. This may be attributed to the fact that both types of variable stars are excited by the $\kappa$ mechanism in the ionization zone of iron-group elements.

These BCEP stars are extremely valuable sources for asteroseismic studies that can significantly enhance our understanding of the theory of the evolution and structure of massive stars. Therefore, it is necessary to study them more thoroughly and identify the specific pulsation modes.



\begin{longtable}{llllllllllll}

\caption{\label{tab:1} The Catalog of BCEP stars observed by TESS and Gaia.}\\
\hline
TESS ID   & $\pi$   & V       & $BC$    & $A_{V}$ & $Teff$   & $log(L_/L_{\odot})$    & Period     & Amplitude & Comments & Flag \\
          & ($mas$) & ($Mag$) & ($Mag$) & ($Mag$) & ($K$)    &                        & ($days$)   & ($mmag$)  &          &      \\

\hline
\endfirsthead

\caption{\label{tab:frequency}(Continued)}\\
\hline
TESS ID   & $\pi$   & V       & $BC$    & $A_{V}$ & $Teff$   & $log(L_/L_{\odot})$    & Period     & Amplitude & Comments & Flag \\
          & ($mas$) & ($Mag$) & ($Mag$) & ($Mag$) & ($K$)    &                        & ($days$)   & ($mmag$)  &          &      \\

\hline \endhead

\hline
\multicolumn{6}{r}{\textsl{(Continued)}}\\
\endfoot

\hline

\endlastfoot

2323229   & 0.40 & 10.32 & -2.33 & 2.17 & 23513(476)  & 4.37 & 0.164521340(1) & 11.85(2)   & EB+BCEP      &        \\
3572937   & 0.34 & 8.86  & -2.56 & 0.99 & 25842(606)  & 4.70 & 0.100887694(1) & 0.47(1)    & BCEP         &        \\
11113493  & 0.47 & 9.90  & -2.41 & 2.46 & 24257(417)  & 4.55 & 0.152174050(1) & 2.54(1)    & BCEP         &        \\
11696250  & 0.48 & 9.46  & -2.51 & 1.85 & 25258(573)  & 4.50 & 0.146738198(1) & 4.06(1)    & BCEP         & 1,2    \\
11698190  & 0.49 & 9.29  & -2.63 & 1.59 & 26539(417)  & 4.48 & 0.170303685(1) & 16.21(1)   & BCEP         & 1,2    \\
12148301  & 0.50 & 10.38 & -2.04 & 1.45 & 20901(407)  & 3.74 & 0.174705217(2) & 1.04(2)    & BCEP         &     3  \\
12249164  & 0.58 & 10.21 & -2.53 & 2.32 & 25469(577)  & 4.22 & 0.307991286(1) & 2.52(1)    & BCEP?(C)     &  (2)   \\
13332837  & 0.60 & 9.80  & -2.30 & 2.12 & 23254(487)  & 4.19 & 0.145467628(1) & 9.14(1)    & BCEP         & 1,2    \\
13967727  & 0.79 & 9.08  & -2.29 & 2.29 & 23094(511)  & 4.31 & 0.225126047(3) & 0.11(1)    & BCEP         & 1,(2)  \\
13972612  & 0.56 & 11.32 & -2.36 & 2.63 & 23782(396)  & 3.86 & 0.094243168(1) & 4.59(1)    & BCEP         &  2     \\
14206699  & 0.51 & 11.88 & -2.59 & 4.19 & 26097(804)  & 4.45 & 0.075896020(1) & 0.35(1)    & BCEP         &  2     \\
22351385  & 0.54 & 11.89 & -2.34 & 2.49 & 23584(754)  & 3.60 & 0.149170187(1) & 55.77(3)   & BCEP(C)      &  2     \\
29036690  & 0.57 & 10.05 & -2.39 & 1.00 & 24086(407)  & 3.72 & 0.126199849(1) & 1.00(1)    & EB+BCEP      & (1)    \\
29585482  &      & 11.31 & -1.94 & 2.23 & 20034(643)  &      & 0.123891425(1) & 6.04(1)    & BCEP         & 1,2,3  \\
40827873  & 0.50 & 10.02 & -2.54 & 1.93 & 25647(554)  & 4.28 & 0.069816023(1) & 0.64(1)    & BCEP         &     3  \\
41327931  & 0.51 & 9.92  & -2.32 & 1.21 & 23408(492)  & 3.93 & 0.111452360(1) & 1.90(1)    & BCEP         & 1,2    \\
42365645  & 0.52 & 9.68  & -2.34 & 1.24 & 23569(580)  & 4.02 & 0.127705997(1) & 26.33(1)   & BCEP         & 1,2    \\
42940133  & 0.77 & 8.49  & -2.25 & 0.91 & 22715(383)  & 3.99 & 0.135368580(1) & 7.83(1)    & BCEP         & 1,2    \\
46408443  & 0.60 & 11.30 & -2.63 & 3.21 & 26603(583)  & 4.16 & 0.145777543(1) & 18.31(2)   & BCEP         &        \\
49217799  & 0.22 & 10.52 & -2.25 & 0.26 & 22708(676)  & 4.02 & 0.092710152(1) & 18.28(1)   & BCEP         &        \\
65516748  & 0.58 & 8.47  & -2.48 & 0.00 & 25013(433)  & 3.98 & 0.116988594(1) & 2.88(1)    & BCEP         &(1),(2) \\
66761070  & 0.26 & 11.43 & -2.90 & 1.86 & 29698(451)  & 4.41 & 0.172448315(1) & 20.78(2)   & BCEP         &        \\
68940684  & 0.27 & 10.20 & -2.57 & 0.42 & 25958(554)  & 4.14 & 0.117758516(1) & 0.74(1)    & BCEP         &        \\
74848209  & 0.30 & 11.48 & -2.69 & 2.88 & 27268(531)  & 4.59 & 0.204847460(1) & 1.02(1)    & BCEP         &        \\
75745359  & 0.92 & 9.45  & -2.10 & 1.88 & 21362(378)  & 3.78 & 0.137758913(1) & 2.86(1)    & BCEP         & 1,2    \\
78205695  & 0.54 & 10.89 & -2.52 & 2.75 & 25354(520)  & 4.18 & 0.185904811(1) & 2.12(1)    & BCEP         &        \\
78652307  & 0.47 & 10.71 & -2.91 & 2.84 & 29781(388)  & 4.57 & 0.174615589(1) & 15.08(1)   & BCEP(C)      &        \\
81400324  & 0.41 & 9.00  & -2.43 & 0.60 & 24515(442)  & 4.28 & 0.097935232(1) & 4.71(1)    & BCEP         &        \\
89756665  & 0.53 & 10.09 & -2.37 & 1.44 & 23899(4345) & 3.94 & 0.118796086(1) & 6.09(1)    & BCEP         & 1,2    \\
93730538  & 0.51 & 9.69  & -2.15 & 1.27 & 21806(537)  & 3.96 & 0.183656256(1) & 9.75(1)    & BCEP(C)      & 1,2,3  \\
105961652 & 0.41 & 11.63 & -2.81 & 3.18 & 28629(384)  & 4.41 & 0.169729856(3) & 22.48(3)   & BCEP         &        \\
106941893 & 0.41 & 9.83  & -2.78 & 2.47 & 28293(446)  & 4.84 & 0.154938412(1) & 5.54(1)    & BCEP         &        \\
108694650 & 0.50 & 10.81 & -2.50 & 2.29 & 25196(518)  & 4.09 & 0.157628505(2) & 20.39(2)   & BCEP         &        \\
113404375 & 0.43 & 10.68 & -2.60 & 2.02 & 26242(942)  & 4.19 & 0.103687693(1) & 16.38(2)   & BCEP         &        \\
121679880 & 0.88 & 9.20  & -2.73 & 2.72 & 27639(283)  & 4.50 & 0.158299626(1) & 5.43(1)    & BCEP         &     3  \\
123211175 & 1.30 & 8.20  & -2.19 & 0.46 & 22159(280)  & 3.44 & 0.139099369(1) & 0.78(1)    & BCEP         & 1,2    \\
125100416 & 0.80 & 9.21  & -2.35 & 1.66 & 23640(439)  & 4.01 & 0.121368126(1) & 4.36(1)    & BCEP         &        \\
129069689 & 0.36 & 11.30 & -2.35 & 2.25 & 23672(651)  & 4.11 & 0.135306124(2) & 20.62(3)   & BCEP         &        \\
131260526 & 0.28 & 10.81 & -2.87 & 2.12 & 29297(716)  & 4.67 & 0.173495855(4) & 50.02(4)   & BCEP(C)      &      3 \\
134522557 & 1.00 & 9.75  & -2.40 & 2.18 & 24157(628)  & 3.83 & 0.174435380(1) & 8.09(1)    & EB+BCEP      &  (2),3 \\
139691604 & 0.23 & 11.13 & -2.64 & 1.79 & 26691(528)  & 4.49 & 0.25565977(2)  & 0.37(4)    & BCEP         &        \\
141903541 & 0.54 & 10.00 & -2.15 & 1.48 & 21842(377)  & 3.89 & 0.189218750(1) & 1.02(1)    & BCEP         &  2     \\
145594454 & 0.52 & 10.21 & -2.03 & 1.66 & 20757(419)  & 3.86 & 0.175192687(1) & 12.58(1)   & BCEP         &        \\
151391621 & 0.35 & 11.86 & -2.85 & 2.59 & 29089(292)  & 4.24 & 0.099888755(1) & 26.40(5)   & BCEP         &        \\
156436102 & 0.56 & 10.81 & -2.68 & 2.44 & 27084(913)  & 4.13 & 0.101212379(1) & 8.10(2)    & BCEP         &        \\
159786058 & 0.63 & 9.68  & -2.34 & 2.11 & 23545(779)  & 4.21 & 0.102341356(1) & 6.04(1)    & BCEP         &      3 \\
168996597 & 0.28 & 9.43  & -2.31 & 0.65 & 23315(380)  & 4.42 & 0.174607903(1) & 3.31(1)    & BCEP         & 1,2    \\
184226481 & 0.41 & 10.27 & -2.45 & 2.05 & 24658(446)  & 4.36 & 0.280963554(1) & 4.02(1)    & BCEP         & (1)    \\
187483462 & 0.49 & 8.78  & -2.49 & 1.28 & 25122(331)  & 4.52 & 0.148482405(1) & 1.08(1)    & BCEP         & 1      \\
189561605 & 0.25 & 11.30 & -2.86 & 2.36 & 29194(550)  & 4.67 & 0.210620111(2) & 7.26(1)    & BCEP         &        \\
190783321 & 1.06 & 8.31  & -2.37 & 1.02 & 23892(448)  & 3.88 & 0.064141626(1) & 1.20(1)    & BCEP         & (1),2  \\
199833554 & 0.50 & 10.27 & -2.43 & 1.99 & 24525(971)  & 4.15 & 0.163292248(1) & 23.57(1)   & BCEP         &        \\
202567458 & 0.71 & 9.87  & -2.89 & 2.45 & 29579(340)  & 4.39 & 0.073436696(1) & 1.79(1)    & BCEP         &  (2),3 \\
214141856 & 0.20 & 11.61 & -2.50 & 1.46 & 25220(762)  & 4.25 & 0.116163042(1) & 17.15(4)   & BCEP(C)      &        \\
215354826 & 0.43 & 11.63 & -2.65 & 3.24 & 26784(585)  & 4.34 & 0.204528308(4) & 18.50(3)   & BCEP         &        \\
216167514 &      & 9.67  & -1.94 & 2.31 & 20000(616)  &      & 0.136146867(1) & 5.69(1)    & BCEP         &      3 \\
217059262 & 0.78 & 9.24  & -2.55 & 1.62 & 25674(571)  & 4.09 & 0.070739471(1) & 2.84(1)    & BCEP         &        \\
229024168 & 0.80 & 9.77  & -2.63 & 1.88 & 26517(581)  & 3.98 & 0.110185875(1) & 11.31(1)   & BCEP         &        \\
234540238 & 0.51 & 10.34 & -2.61 & 1.86 & 26388(835)  & 4.14 & 0.217702497(4) & 2.09(2)    & BCEP         &        \\
235416305 & 0.40 & 10.64 & -2.55 & 1.78 & 25744(805)  & 4.17 & 0.168833412(2) & 14.97(2)   & BCEP         &        \\
235518467 & 0.39 & 10.82 & -2.75 & 1.99 & 27893(701)  & 4.29 & 0.166051737(1) & 23.42(2)   & BCEP         &  2     \\
236073671 & 0.40 & 10.22 & -2.64 & 2.09 & 26657(375)  & 4.50 & 0.228633692(3) & 34.48(1)   & BCEP         &        \\
245496621 & 0.39 & 10.84 & -2.46 & 2.58 & 24817(763)  & 4.39 & 0.135203437(1) & 12.18(2)   & BCEP(C)      &        \\
245719692 & 0.45 & 10.69 & -2.41 & 2.37 & 24229(684)  & 4.22 & 0.144377151(1) & 2.66(2)    & BCEP         & 1,2    \\
246785266 &      & 10.66 & -2.47 & 3.04 & 24921(1011) &      & 0.093001342(1) & 13.41(1)   & BCEP(C)      &        \\
246954609 & 0.67 & 9.81  & -2.67 & 1.75 & 27035(611)  & 4.09 & 0.066646826(1) & 1.81(1)    & BCEP         &        \\
247649462 & 0.84 & 9.37  & -2.21 & 1.84 & 22405(374)  & 3.92 & 0.184293586(2) & 0.13(2)    & BCEP         &        \\
251196433 & 0.37 & 10.12 & -2.65 & 2.26 & 26792(523)  & 4.67 & 0.156885338(1) & 8.65(1)    & BCEP         & 1,(2)  \\
251250634 & 0.57 & 10.71 & -2.44 & 2.42 & 24617(621)  & 4.05 & 0.075731665(1) & 2.56(2)    & BCEP         & 1,2    \\
253360159 & 0.55 & 9.33  & -2.84 & 1.41 & 28908(232)  & 4.38 & 0.101012661(1) & 1.57(1)    & BCEP         &        \\
254365178 & 0.28 & 10.73 & -2.63 & 0.82 & 26530(802)  & 4.08 & 0.115963137(1) & 11.79(2)   & BCEP         &        \\
254684802 & 0.21 & 10.71 & -1.94 & 1.16 & 20000(321)  & 4.22 & 0.156966337(2) & 18.65(2)   & BCEP         &  2     \\
255974332 & 0.45 & 10.99 & -2.39 & 2.90 & 24058(451)  & 4.32 & 0.204341179(3) & 19.73(2)   & BCEP         & 1,2    \\
262639816 & 0.45 & 11.90 & -2.38 & 3.17 & 23958(1229) & 4.05 & 0.152539053(3) & 35.71(4)   & BCEP         &     3  \\
264615882 & 0.50 & 8.86  & -2.67 & 1.19 & 26955(265)  & 4.50 & 0.2124214(6)   & 39(3)      & BCEP         & 1,2    \\
266338052 & 0.53 & 8.92  & -2.18 & 1.94 & 22125(272)  & 4.52 & 0.174225800(1) & 7.64(1)    & BCEP         & 1   3  \\
269228628 & 0.33 & 10.90 & -2.80 & 2.18 & 28440(411)  & 4.50 & 0.20060282(1)  & 0.74(6)    & EB+BCEP(C)   & 1,2,3  \\
271610589 & 0.30 & 9.89  & -2.44 & 1.40 & 24527(522)  & 4.52 & 0.257073465(8) & 0.37(2)    & BCEP         &     3  \\
272922818 & 0.32 & 10.90 & -2.88 & 2.69 & 29396(387)  & 4.75 & 0.124569258(1) & 12.40(1)   & BCEP         & 1,2    \\
274070426 & 0.50 & 10.06 & -2.63 & 2.70 & 26537(373)  & 4.62 & 0.228697289(1) & 0.45(1)    & BCEP         &        \\
279659875 & 0.94 & 9.14  & -2.11 & 2.09 & 21501(335)  & 3.97 & 0.171524091(1) & 5.56(1)    & BCEP         & 1,(2),3\\
282529703 & 0.49 & 10.42 & -2.10 & 1.20 & 21366(227)  & 3.67 & 0.133324088(1) & 10.93(1)   & BCEP         & 1      \\
287690192 & 0.81 & 9.69  & -2.32 & 1.84 & 23368(423)  & 3.87 & 0.120925520(1) & 10.76(1)   & BCEP         & 1,2    \\
292765671 & 0.59 & 10.47 & -2.57 & 2.45 & 25952(510)  & 4.18 & 0.162282280(1) & 18.97(1)   & BCEP         &        \\
294799849 & 0.51 & 9.89  & -2.27 & 1.07 & 22889(489)  & 3.86 & 0.086501321(1) & 2.13(1)    & BCEP(C)      &  2,3   \\
296090437 & 0.34 & 11.10 & -2.35 & 1.58 & 23668(448)  & 3.97 & 0.278424790(1) & 1.08(1)    & BCEP(C)      &    3   \\
296570221 & 0.39 & 10.53 & -2.31 & 1.48 & 23343(487)  & 4.01 & 0.164183036(1) & 12.68(1)   & BCEP(C)      & 1,2    \\
297259536 & 1.24 & 8.49  & -2.62 & 2.46 & 26402(312)  & 4.34 & 0.087836186(1) & 2.52(1)    & BCEP         & 1,2    \\
299962330 & 0.50 & 10.67 & -2.28 & 1.14 & 22972(331)  & 3.60 & 0.121930769(1) & 9.68(2)    & BCEP         &        \\
301779501 & 0.54 & 9.09  & -2.35 & 0.77 & 23703(453)  & 4.05 & 0.218233499(3) & 17.19(2)   & BCEP(C)      &        \\
303136029 & 0.24 & 10.40 & -2.56 & 0.63 & 25790(421)  & 4.24 & 0.133134490(1) & 27.20(1)   & BCEP         &        \\
304958470 & 0.42 & 9.91  & -2.15 & 0.96 & 21831(423)  & 3.92 & 0.183697184(1) & 5.28(1)    & BCEP(C)      &        \\
306386296 & 0.38 & 9.59  & -2.31 & 1.02 & 23257(471)  & 4.23 & 0.207592140(3) & 1.95(2)    & BCEP(C)      &  2     \\
308339884 & 0.39 & 9.14  & -2.52 & 0.69 & 25445(192)  & 4.35 & 0.303891059(4) & 4.18(2)    & BCEP?        &     3  \\
308954763 & 0.46 & 9.26  & -2.29 & 0.62 & 23105(455)  & 4.02 & 0.195026416(1) & 1.57(2)    & BCEP         & 1,2    \\
311943795 & 0.32 & 9.97  & -2.68 & 1.46 & 27063(664)  & 4.56 & 0.120640928(1) & 5.57(1)    & BCEP         & 1,2,3  \\
312626970 & 0.53 & 9.56  & -2.50 & 1.93 & 25187(369)  & 4.40 & 0.219434891(1) & 1.79(2)    & BCEP         & 1,2    \\
312630206 & 0.53 & 9.60  & -2.40 & 1.54 & 24216(676)  & 4.19 & 0.164352021(1) & 25.73(1)   & BCEP         & 1,2    \\
312637783 & 0.52 & 10.20 & -2.23 & 1.50 & 22561(382)  & 3.88 & 0.122231162(1) & 7.27(1)    & BCEP         & 1,(2)  \\
312639883 & 0.49 & 9.25  & -2.50 & 1.32 & 25222(529)  & 4.35 & 0.167984487(1) & 6.61(1)    & BCEP(C)      & 1,2    \\
314393072 & 0.38 & 10.27 & -2.45 & 1.38 & 24720(432)  & 4.17 & 0.239163295(1) & 1.79(1)    & BCEP         &        \\
314529804 & 0.41 & 10.25 & -2.32 & 1.81 & 23381(503)  & 4.22 & 0.169392516(1) & 22.51(1)   & BCEP(C)      &  2,3   \\
315506577 & 0.30 & 10.79 & -2.07 & 1.59 & 21147(580)  & 4.08 & 0.218518296(1) & 11.34(1)   & BCEP(C)      &        \\
317053810 & 0.42 & 9.94  & -2.30 & 0.94 & 23170(425)  & 3.96 & 0.166332341(1) & 10.28(1)   & BCEP         &  (2)   \\
318667559 & 0.47 & 10.15 & -2.43 & 0.98 & 24453(449)  & 3.85 & 0.067096349(1) & 2.94(1)    & BCEP(C)      &      3 \\
322419186 & 0.40 & 9.71  & -2.38 & 0.54 & 23943(361)  & 3.97 & 0.157841597(1) & 1.61(1)    & BCEP         &  2     \\
328165300 & 0.25 & 11.87 & -2.29 & 1.76 & 23157(429)  & 3.97 & 0.137560593(1) & 11.56(3)   & BCEP         &  2     \\
336932754 & 0.31 & 11.34 & -2.61 & 1.78 & 26386(518)  & 4.14 & 0.115250943(1) & 10.53(4)   & BCEP         &      3 \\
337028799 & 0.27 & 10.68 & -2.36 & 1.38 & 23739(539)  & 4.26 & 0.168518408(1) & 25.49(1)   & BCEP         &        \\
337105253 & 0.96 & 10.43 & -2.22 & 3.28 & 22442(329)  & 3.96 & 0.168178145(1) & 5.75(1)    & BCEP         & 1,(2),3\\
338637985 & 0.61 & 9.91  & -2.40 & 1.21 & 24183(525)  & 3.81 & 0.069368003(1) & 12.35(1)   & BCEP         &  2     \\
338681866 & 0.46 & 10.63 & -2.48 & 1.88 & 25007(577)  & 4.07 & 0.160341609(1) & 15.11(1)   & BCEP         &      3 \\
338733059 & 0.60 & 10.58 & -2.20 & 1.86 & 22256(489)  & 3.73 & 0.100673530(1) & 7.08(1)    & BCEP         &        \\
348137274 & 0.37 & 9.33  & -2.14 & 1.20 & 21712(394)  & 4.37 & 0.157540717(1) & 7.33(1)    & BCEP         & 1,2    \\
352229000 & 0.37 & 10.23 & -1.95 & 0.18 & 20131(410)  & 3.51 & 0.087022219(1) & 11.49(1)   & BCEP         &  2,3   \\
360062316 & 0.44 & 11.48 & -2.65 & 2.38 & 26748(330)  & 4.03 & 0.153438623(1) & 0.34(2)    & BCEP?        &      3 \\
360079950 & 1.46 & 10.90 & -2.62 & 2.11 & 26424(321)  & 3.10 & 0.059847319(1) & 0.72(2)    & BCEP         &        \\
360371656 & 0.28 & 10.09 & -2.74 & 0.96 & 27740(413)  & 4.45 & 0.213397967(4) & 0.19(2)    & BCEP(C)      &  2     \\
360981544 & 0.13 & 11.99 & -2.25 & 1.73 & 22776(738)  & 4.45 & 0.230261335(2) & 0.95(2)    & BCEP         &        \\
361324132 & 0.48 & 9.61  & -2.34 & 1.67 & 23580(317)  & 4.29 & 0.152251019(1) & 13.33(1)   & BCEP         & 1,2    \\
363706224 & 0.10 & 11.84 & -2.23 & 0.54 & 22596(479)  & 4.25 & 0.224945407(1) & 15.19(1)   & BCEP         &        \\
366108295 & 0.39 & 10.20 & -2.54 & 1.26 & 25625(493)  & 4.16 & 0.077180697(1) & 3.60(1)    & BCEP         & (1)    \\
372674359 & 0.20 & 10.70 & -1.94 & 0.66 & 20000(109)  & 4.05 & 0.152802348(1) & 29.61(1)   & BCEP         &  2,3   \\
378429048 & 0.77 & 9.78  & -2.14 & 1.91 & 21698(382)  & 3.83 & 0.153248346(1) & 8.21(1)    & BCEP         & 1,2,3  \\
379932247 & 0.20 & 11.27 & -2.08 & 0.72 & 21192(281)  & 3.91 & 0.107457275(1) & 3.77(3)    & BCEP(C)      & (1),2  \\
380435021 & 0.59 & 10.97 & -2.54 & 2.69 & 25595(680)  & 4.05 & 0.117202885(1) & 18.32(2)   & BCEP         &        \\
380870688 & 0.93 & 9.64  & -2.28 & 1.94 & 23025(423)  & 3.79 & 0.107018347(1) & 4.72(1)    & EB+BCEP      &        \\
383228890 & 0.27 & 11.48 & -2.86 & 2.34 & 29160(302)  & 4.53 & 0.226814850(1) & 0.81(1)    & BCEP         &        \\
384540878 & 0.12 & 11.75 & -2.87 & 1.01 & 29304(372)  & 4.62 & 0.290094127(1) & 3.21(1)    & BCEP         & (1)    \\
385457089 & 0.47 & 9.89  & -2.47 & 0.55 & 24897(438)  & 3.80 & 0.128985652(1) & 0.94(1)    & BCEP         &        \\
391836737 & 0.87 & 8.39  & -2.26 & 0.95 & 22787(362)  & 3.94 & 0.161493087(1) & 14.11(1)   & BCEP         & 1      \\
395218466 & 0.16 & 11.50 & -2.45 & 2.51 & 24630(589)  & 4.85 & 0.274682478(1) & 2.92(1)    & EB+BCEP      &        \\
397221078 & 0.38 & 9.55  & -2.53 & 1.64 & 25519(527)  & 4.58 & 0.160588496(1) & 6.58(1)    & BCEP         &  2     \\
400852783 & 0.51 & 10.99 & -2.80 & 2.80 & 28479(929)  & 4.33 & 0.145343978(1) & 9.59(1)    & BCEP         & 1,2    \\
410100465 & 0.53 & 10.76 & -2.40 & 2.37 & 24189(385)  & 4.05 & 0.130122864(1) & 4.08(2)    & BCEP(C)      &        \\
410672776 & 0.40 & 9.66  & -2.83 & 1.88 & 28839(267)  & 4.71 & 0.175321034(1) & 2.55(1)    & BCEP         &   3    \\
414281010 & 0.27 & 10.48 & -2.28 & 0.83 & 23004(488)  & 4.08 & 0.164426934(1) & 4.09(1)    & BCEP         &  2,3   \\
420545430 & 0.27 & 11.03 & -2.66 & 2.87 & 26861(490)  & 4.84 & 0.304953378(1) & 4.33(1)    & BCEP?        & (1)    \\
421117635 & 0.33 & 10.55 & -2.67 & 2.75 & 27042(468)  & 4.80 & 0.192028939(1) & 5.83(1)    & BCEP         & 1      \\
428380374 & 0.48 & 9.48  & -2.43 & 2.34 & 24466(418)  & 4.64 & 0.201759588(5) & 3.71(3)    & BCEP         &  2     \\
430460842 & 0.26 & 9.92  & -2.43 & 1.22 & 24441(512)  & 4.55 & 0.210530997(6) & 31.46(3)   & BCEP         &  2,3   \\
430625041 & 0.75 & 9.08  & -2.03 & 0.92 & 20808(407)  & 3.69 & 0.173222541(1) & 8.93(1)    & BCEP         & (1),3  \\
434362297 & 0.47 & 11.13 & -2.48 & 3.22 & 24975(415)  & 4.39 & 0.200198922(1) & 2.82(1)    & BCEP         &     3  \\
439575025 & 0.47 & 10.44 & -2.10 & 1.15 & 21398(394)  & 3.68 & 0.154382233(1) & 10.19(2)   & BCEP         &        \\
445380881 & 0.51 & 8.14  & -2.29 & 0.74 & 23130(210)  & 4.44 & 0.140483103(1) & 2.09(1)    & BCEP         &  2     \\
446524355 & 0.28 & 10.50 & -2.28 & 1.02 & 23043(451)  & 4.11 & 0.22450480(4)  & 3.0(4)     & BCEP         &        \\
449526185 & 0.53 & 9.13  & -2.45 & 1.33 & 24705(403)  & 4.31 & 0.171648931(1) & 9.96(1)    & BCEP         &  2     \\
452018494 & 0.54 & 9.92  & -2.16 & 1.37 & 21952(417)  & 3.87 & 0.115552570(1) & 5.56(1)    & BCEP(C)      & 1,2    \\
458263480 & 0.44 & 10.30 & -2.01 & 1.30 & 20572(373)  & 3.82 & 0.174223657(1) & 8.65(1)    & EB+BCEP      &        \\
458704068 & 0.53 & 9.55  & -2.50 & 0.86 & 25204(537)  & 3.97 & 0.154750772(1) & 6.39(1)    & BCEP         &  2     \\
460719374 & 0.60 & 10.32 & -2.33 & 1.91 & 23511(455)  & 3.91 & 0.134054696(1) & 4.87(1)    & BCEP         &  2     \\
467023914 & 0.45 & 10.10 & -2.22 & 0.95 & 22438(445)  & 3.81 & 0.148579638(1) & 11.62(1)   & BCEP(C)      &  2     \\
467448794 & 0.45 & 10.30 & -2.35 & 2.09 & 23675(476)  & 4.25 & 0.185468072(2) & 2.09(1)    & BCEP         &  (2)   \\
467480241 & 0.37 & 9.71  & -1.90 & 0.89 & 19719(144)  & 4.00 & 0.142144653(1) & 6.75(1)    & BCEP         &        \\
469223889 & 0.49 & 9.21  & -2.43 & 0.52 & 24428(467)  & 4.01 & 0.127504565(1) & 4.82(1)    & BCEP(C)      & 1,2    \\

\end{longtable}
{Note. The numbers in the parentheses in Columns 6, 8, and 9 represent the errors of the data. The capital letter "C" in the parentheses of Column 10 represents targets that may be contaminated by neighboring stars in the TESS photometry apertures. Some targets that are also listed in other literature are given in Column 11 as follows: 1 - listed as BCEP stars by \citet{2020AJ....160...32L}; (1) - listed as candidates by \citet{2020AJ....160...32L}; 2 - listed as BCEP stars by \citet{2020MNRAS.493.5871B}; (2) - listed in the Catalog of \citet{2020MNRAS.493.5871B}, but not as BCEP stars; 3 - listed as MS oscillators by Gaia DR3.\\}

\begin{acknowledgments}

This work is partly supported by the International Cooperation Projects of the National Key R\&D Program (No. 2022YFE0127300), Chinese Natural Science Foundation (Nos. 11933008, 12103084, and 12273103), Yunnan Revitalization Talent Support Program, and the basic research project of Yunnan Province (Grant No. 202301AT070352).
This work has made use of data from the European Space Agency (ESA) mission Gaia (https://www.cosmos.esa.int/gaia), processed by the Gaia Data Processing and Analysis Consortium (DPAC, https://www.cosmos.esa.int/web/gaia/dpac/consortium). Funding for the DPAC has been provided by national institutions, in particular the institutions
participating in the Gaia Multilateral Agreement.
The TESS data presented in this paper were obtained from the Mikulski Archive for Space Telescopes (MAST) at the Space Telescope Science Institute (STScI). STScI is operated
by the Association of Universities for Research in Astronomy, Inc. Support to MAST for these data is provided by the NASA Office of Space Science. Funding for the TESS mission is provided by the NASA Explorer Program.

\end{acknowledgments}

\appendix

\section{Fourier spectra}

\begin{figure*}\centering \vbox to9.5in{\rule{0pt}{5.0in}}
\includegraphics{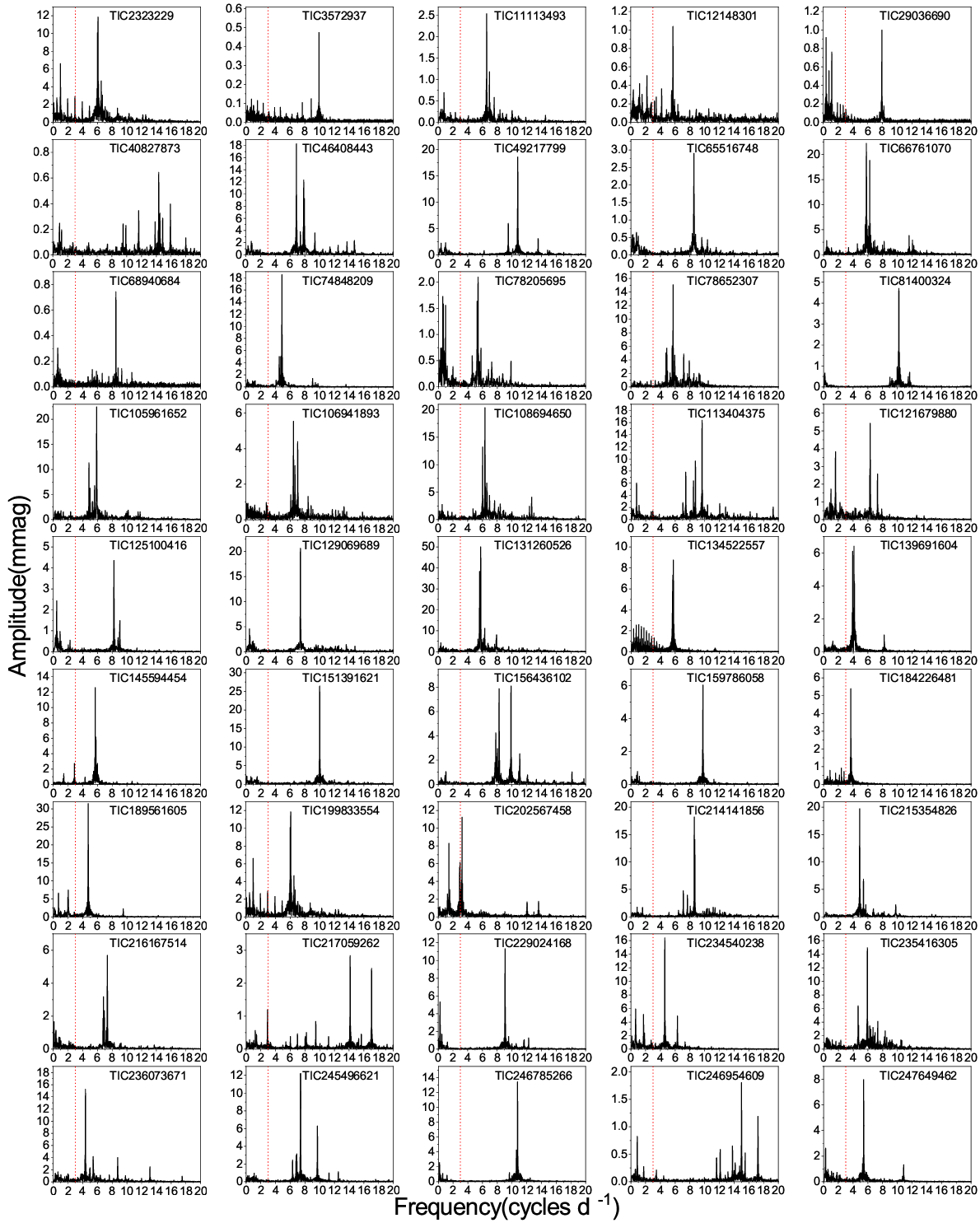}
\caption{Fourier spectra of 83 stars first confirmed as BCEP stars, part 1.}
\label{fig:FS1}
\end{figure*}

\begin{figure*}\centering \vbox to9.5in{\rule{0pt}{5.0in}}
\includegraphics{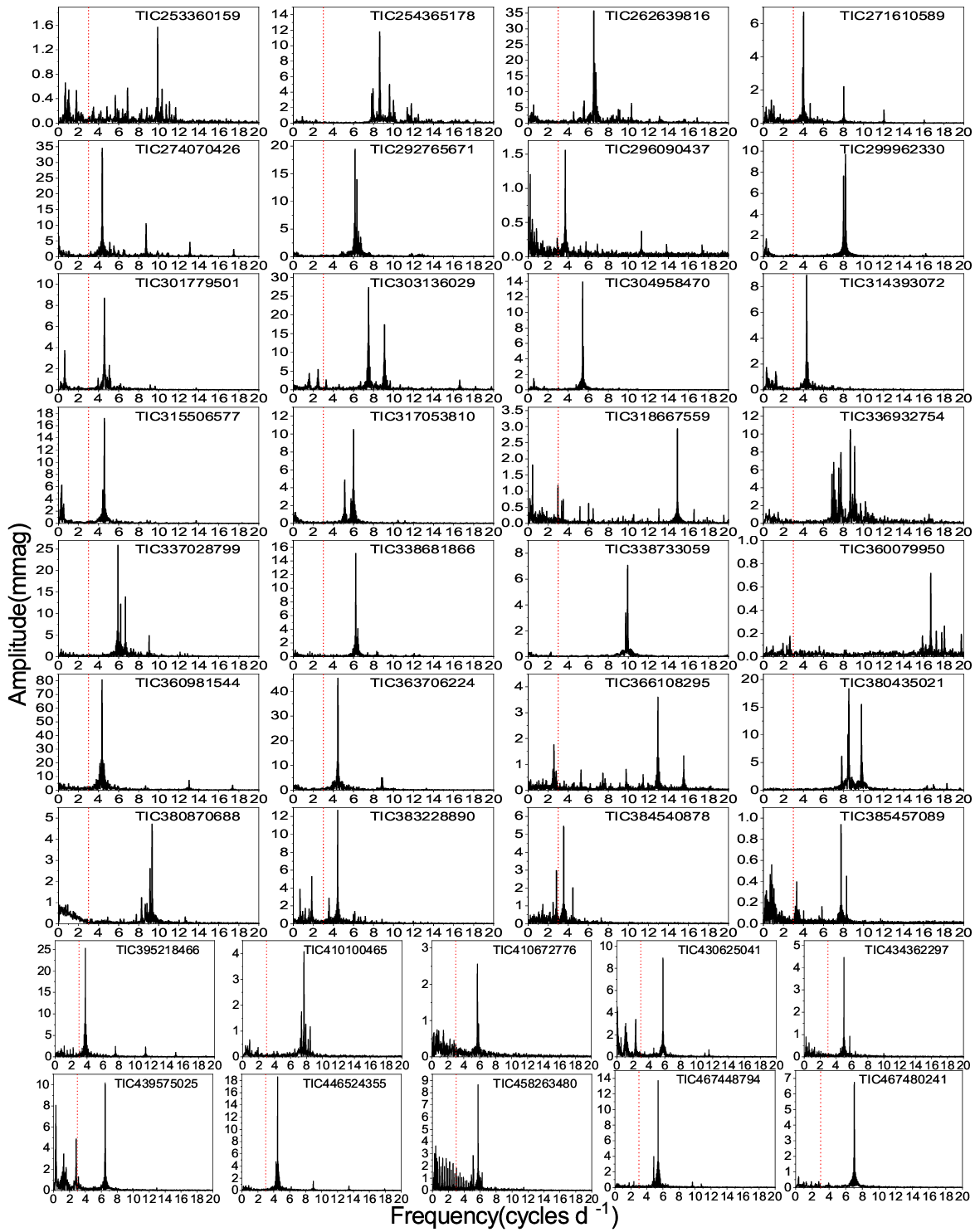}
\caption{Fourier spectra of 83 stars first confirmed as BCEP stars, part 2.}
\label{fig:FS2}
\end{figure*}

The Fourier spectra of 83 stars first confirmed as BCEP stars are shown in Figures \ref{fig:FS1} and \ref{fig:FS2}.

\end{document}